\documentclass[review]{elsarticle}

\usepackage{lineno,hyperref}
\usepackage{graphicx}
\usepackage{float}
\usepackage{xcolor}
\usepackage{soul}
\usepackage{amsmath}
\usepackage{subcaption}
\usepackage{multirow}
\usepackage{lscape}
\usepackage{bm}
\usepackage{tabularx,booktabs}
\usepackage{arydshln}

\newcommand{\ie}{i.\,e.}

\modulolinenumbers[5]

\newcolumntype{Y}{>{\centering\arraybackslash}X}

\begin{document}

\begin{frontmatter}

\title{Fitbeat: COVID-19 Estimation based on Wristband Heart Rate}


\author[add1]{Shuo Liu\corref{mycorrespondingauthor}}
\ead{shuo.liu@informatik.uni-augsburg.com}
\author[add1,add2]{Jing Han}
\author[add3,add4]{Estela Laporta Puyal}
\author[add3,add4]{Spyridon Kontaxis}
\author[add5]{Shaoxiong Sun}
\author[add6]{Patrick Locatelli}
\author[add1]{Judith Dineley}
\author[add1,add7]{Florian B.\ Pokorny}
\author[add8]{Gloria Dalla Costa}
\author[add8]{Letizia Leocani}
\author[add9]{Ana Isabel Guerrero}
\author[add9]{Carlos Nos}
\author[add9]{Ana Zabalza}
\author[add10]{Per Soelberg S\o{}rensen}
\author[add10]{Mathias Buron}
\author[add10]{Melinda Magyari}
\author[add5]{Yatharth Ranjan}
\author[add5]{Zulqarnain Rashid}
\author[add5]{Pauline Conde}
\author[add5]{Callum Stewart}
\author[add5,add11]{Amos A Folarin}
\author[add5,add11]{Richard JB Dobson}
\author[add3,add4]{Raquel Bail\'on}
\author[add12]{Srinivasan Vairavan}
\author[add1,add5]{Nicholas Cummins}
\author[add12]{Vaibhav A Narayan}
\author[add13,add14]{Matthew Hotopf}
\author[add15]{Giancarlo Comi}
\author[add1,add16]{Bj\"orn Schuller}
\author[add17]{RADAR-CNS Consortium}

\address[add1]{EIHW -- Chair of Embedded Intelligence for Health Care and Wellbeing, University of Augsburg, Augsburg, Germany}
\address[add2]{Department of Computer Science and Technology, University of Cambridge, Cambridge, United Kingdom}
\address[add3]{BSICoS Group, Arag\'on Institute of Engineering Research (I3A), IIS Arag\'on, University of Zaragoza, Zaragoza, Spain}
\address[add4]{CIBER of Bioengineering, Biomaterials and Nanomedicine (CIBER-BNN), Madrid, Spain}
\address[add5]{The Department of Biostatistics and Health informatics, Institute of Psychiatry, Psychology and Neuroscience, King’s College London, London, UK}
\address[add6]{Department of Engineering and Applied Science, University of Bergamo, Bergamo, Italy}
\address[add7]{Division of Phoniatrics, Medical University of Graz, Graz, Austria}
\address[add8]{Neurorehabilitation Unit and Institute of Experimental Neurology, University Vita Salute San Raffaele, Istituto Di Ricovero e Cura a Carattere Scientifico Ospedale San Raffaele, Milan, Italy}
\address[add9]{Multiple Sclerosis Centre of Catalonia (Cemcat), Department of Neurology\/Neuroimmunology, Hospital Universitari Vall d\'Hebron, Universitat Autònoma de Barcelona, Barcelona, Spain}
\address[add10]{Danish Multiple Sclerosis Centre, Department of Neurology, Copenhagen University Hospital Rigshospitalet, Copenhagen, Denmark}
\address[add11]{Institute of Health Informatics, University College London, London, United Kingdom}
\address[add12]{Janssen Research and Development LLC, Titusville, NJ, USA}
\address[add13]{The Department of Psychological Medicine, Institute of Psychiatry, Psychology and Neuroscience, King’s College London, London, United Kingdom}
\address[add14]{South London and Maudsley National Health Service Foundation Trust, London, United Kingdom}
\address[add15]{Institute of Experimental Neurology, Istituto Di Ricovero e Cura a Carattere Scientifico Ospedale San Raffaele, Milan, Italy}
\address[add16]{GLAM -- Group on Language, Audio, \& Music, Imperial College London, London, United Kingdom}
\address[add17]{The RADAR-CNS Consortium, London, United Kingdom. www.radar-cns.org}

\cortext[mycorrespondingauthor]{Corresponding author}

\newpage
\begin{abstract}
This study investigates the potential of deep learning methods to identify individuals with suspected COVID-19 infection using remotely collected heart-rate data. The study utilises data from the ongoing EU IMI RADAR-CNS research project that is investigating the feasibility of wearable devices and smartphones to monitor individuals with multiple sclerosis (MS), depression or epilepsy. As part of the project protocol, heart-rate data was collected from participants using a Fitbit wristband. The presence of COVID-19 in the cohort in this work was either confirmed through a positive swab test, or inferred through the self-reporting of a combination of symptoms including fever, respiratory symptoms, loss of smell or taste, tiredness and gastrointestinal symptoms. 
Experimental results indicate that our proposed contrastive convolutional auto-encoder (contrastive CAE), i.\,e., a combined architecture of an auto-encoder and contrastive loss, outperforms a conventional convolutional neural network (CNN), as well as a convolutional auto-encoder (CAE) without using contrastive loss. 
Our final contrastive CAE achieves $95.3\%$ unweighted average recall, $86.4\%$ precision, an F1 measure of $88.2\%$, a sensitivity of $100\%$ and a specificity of $90.6\%$ on a test set of $19$ participants with MS who reported symptoms of COVID-19. Each of these participants was paired with a participant with MS with no COVID-19 symptoms. 

\end{abstract}

\begin{keyword}
COVID-19\sep respiratory tract infection\sep anomaly detection\sep contrastive learning\sep convolutional auto-encoder\sep deep learning
\end{keyword}

\end{frontmatter}


\section{Introduction}

Remote, passive monitoring of physiological and behavioural characteristics using smartphones and wearable devices can be used to rapidly collect a variety of data in huge volumes. Such data has the potential to improve our understanding of the interplay between a variety of health conditions at both the level of the individual and the population, if rigorously collected and validated \cite{Piwek2016}.

Passive data collection is typically implemented with a high temporal resolution \cite{Ranjan2019}. Wearable fitness trackers, for example, estimate parameters such as heart rate up to every second, and can be worn 24 hours a day. Monitoring individuals with a range of health states, lifestyles, and demographic variables in combination with data artefacts and missing data leads to high variability. From heart rate and general physical activity to GPS-based location information or the proximity to other individuals detected through Bluetooth, multiple types of data can be monitored remotely. Therefore, studies using wearables and smartphones in this way exhibit several ‘v’s of big data, namely velocity, volume, variability and variety. As such, advanced analysis methodology such as deep learning have the potential to make a significant contribution \cite{Mohammadi2018}.  

The volume and richness of remotely collected data has particular potential in the context of infectious diseases, such as the novel corona virus (SARS-CoV-2). Specific applications include individual screening and population-level monitoring while minimising contact with infected individuals \cite{Radin2020, nature2020, zhu2020}.

Among all remotely collectable physiological data, heart rate is a biomarker of particular value in such applications. Patterns in heart rate fluctuations over time  in particular have been 
found to provide clinically relevant information about the integrity of the physiological system generating these dynamics. Previous studies have not only revealed an altered heart rate variability in a number of medical conditions, such as hypovolaemia \cite{Triedman1993}, heart failure \cite{Bonaduce1999}, or angina \cite{Huang1995}, but also demonstrated that the degree of short-term heart rate alteration correlates with illness severity. Analysis of the autonomic regulation of heart rate, such as heart rate variability, has also been discussed as a promising approach for detecting infections earlier than conventional clinical methods and prognosticating infections \cite{Ahmad2009}.

Wearables such as Fitbit fitness trackers\footnote{https://www.fitbit.com/ [as of 12 November 2020]} can provide indirect measurements of the heart rate through pulse rate estimates made using photoplethysmography. Radin and colleagues \cite{Radin2020} analysed the resting heart rate collected in this way, alongside with sleep duration data collected from over 47\,000 individuals to improve model predictions of influenza rates in five US states. In the ongoing DETECT\footnote{http://detectstudy.org/ [as of 12 November 2020]} study\,\cite{nature2020}, the same group is now applying this approach to monitor outbreaks of viral infections including COVID-19. In separate work, Fitbit devices have also been used to identify individuals with COVID-19 using heart rate measurements collected before symptom onset \cite{Mishra2020}.  
In addition to pre-print studies \cite{Marinsek2020, Natarajan2020}, other similar ongoing endeavours include the German project \textit{Corona-Datenspende}\footnote{http://corona-datenspende.de/science/en/ [as of 12 Nov 2020]}, which has a cohort of over 500\,000 volunteers, and the TemPredict study in the US\footnote{http://osher.ucsf.edu/research/current-research-studies/tempredict [as of 12 November 2020]}.

Applied to such data sets, deep learning methodology has the potential to automatically identify individuals with COVID-19 purely on the basis of passive data from wearable devices\,\cite{nature2020,Natarajan2020,Mishra2020}. Natarajan and colleagues \cite{Natarajan2020} used a convolutional neural network (CNN) to predict illness on a given day using Fitbit data from 1\,181 individuals, reporting an area under the receiver operating characteristics curve (AUC) of $0.77\pm0.03$. Similarly, this paper describes a CNN-based deep learning approach for predicting the presence of COVID-like symptoms using Fitbit heart rate data. Treating this task as analogous to anomaly detection, we explore the suitability of using a convolutional auto-encoder (CAE) with contrastive loss \cite{Aytekin2018, Chen2018}. The technique's performance is compared against a conventional auto-encoder and a convolutional neural network. 

\section{Data Collection}
The data used in this work was collected as part of the IMI2 RADAR-CNS major programme\footnote{\url{https://www.radar-cns.org/}}. RADAR-CNS is an observational, longitudinal, prospective study aimed at developing new ways of measuring major depressive disorder, epilepsy, and multiple sclerosis (MS) using remote monitoring technologies (RMT), \ie, wearable devices and smartphone technology. The study is currently being conducted at multiple clinical sites spread across several European countries. The data used in this work was collected in the MS arm of the project with participants recruited from three sites: Ospedale San Raffaele (OSR) in Milan, Italy, Vall d'Hebron Institut de Recerca (VHIR) in Barcelona, Spain, and the University Hospital Copenhagen, Regshospitalet (RegionH) in Copenhagen, Denmark. The MS work package started recruiting participants in June 2018. As of  March 1, 2020, 499 participants had been enrolled in the RADAR-MS study and 403 (81\,\%) remained in the study. For further information on the MS study and the inclusion criteria and assessment schemes the interested reader is referred to~\cite{DallaCosta2020}.

RADAR-CNS collects both passive and active participant data. Passive data is collected on a continuous 24/7 basis through a smartphone and wearable device in a manner that requires minimal participant effort. It includes GPS-derived location, Bluetooth, phone usage data, activity, sleep, and heart rate. Participants use their own Android smartphones where available, or were provided with a Motorola G5, G6, or G7 if required. Fitbit Charge 2 or Charge 3 devices were provided to participants, who were asked to wear the device on their non-dominant hand. 
The collection of active data requires direct input from the participant and includes surveys, questionnaires, and short tasks such as speech recordings. Data collection and management were handled by RADAR-BASE\footnote{\url{https://radar-base.org/}}, a purpose built, open-source mobile Health (m-Health) technology platform~\cite{Ranjan2019}. To assess the impact of COVID-19 on the MS study, a specially designed active questionnaire was distributed, via RADAR-base 
to all active participants first on on March 25, 2020, and then again on April 8, 2020. By April 15 2020, at least one of the questionnaires had been  completed by 399 participants (99\,\%). At the time of these questionnaires, many confirmed COVID-19 cases did not fulfil the criteria as defined by the World Health Organization (WHO)~\cite{huang2020clinical}. The RADAR-CNS MS workpackage used two alternative definitions to determine the prevalence of COVID-19 in their participants~\cite{DallaCosta2020}. The first case definition, herein referred to as CD1, was participants experiencing fever or anosmia/ageusia in combination with any other COVID-19 symptoms including respiratory symptoms, tiredness and gastrointestinal symptoms
, or respiratory symptoms plus two other COVID-19 symptoms, with the second case definition, CD2, being participants experiencing fever + any other COVID-19 symptoms, or respiratory symptoms + anosmia/ageusia. Laboratory-confirmed cases were included in both case definitions~\cite{DallaCosta2020}.

\begin{table}[th]
\caption{Gender-, age-, and site-related distribution of participants per data subset}
\label{tab:participants}
  \centering
  \setlength{\tabcolsep}{0.6mm}
  \setlength{\arrayrulewidth}{0.6pt}
  \renewcommand{\arraystretch}{0.9}
\begin{tabular}{l|l|ccc}
 \multicolumn{1}{c}{} & \multicolumn{1}{c}{} & \multicolumn{1}{c}{} & \textbf{Positive participants} & \textbf{Health Control} \\
 \multicolumn{1}{c}{} & \multicolumn{1}{c}{} & \textbf{Pre-training} & \textbf{for testing} & \textbf{for testing} \\
 \hline
 \multirow{2}{*}{\textbf{Genders}} & \textbf{Female} &  $14$ & $5$ & $5$ \\
                                   & \textbf{Male}   &  $35$ & $14$ & $14$ \\
 \hline
 \multirow{3}{*}{\textbf{Locations}} & \textbf{Italy}  & $18$ &  $7$ &  $7$  \\
 &  \textbf{Spain}         &    $19$      &      $6$ & $6$      \\
 &  \textbf{Denmark}         &    $12$     &     $6$ & $6$         \\
 \hline
 \multirow{6}{*}{\textbf{Ages}} &   $\bm{\leq 30}$   &     $1$  & $2$ & $2$  \\
 &   \textbf{30 - 39}        &    $10$      &       $3$ & $4$         \\
 &   \textbf{40 - 49}        &  $12$        &       $6$ & $5$              \\
 &   \textbf{50 - 59}         &     $19$     &      $6$ & $6$          \\
 &   \textbf{60 - 69}          &    $6$      &      $1$ & $1$          \\
 &   $\bm{\geq 70}$         &   $1$       &          $-$    & $-$   \\
 \hline
\end{tabular}
\end{table}


For this study, we considered Fitbit heart rate measurements, collected 24-hours-a-day/7-days-a-week between 21 February and 20 May 2020, from $87$ participants in Denmark, Italy and Spain, with an age range from 23 to 73 years (mean = $46.5 \pm 10.5$ standard deviation). Therein, 68 MS participants (30 female, 38 male) reported to have symptoms characteristic of COVID-19. However, in 49, symptoms did not meet CD1 or CD2 criteria. The heart rate data of these $49$ participants were used for model pre-training (pre-training set). For testing, we applied leave one subject out (LOSO) cross-validation (CV)
\cite{wong2015} on the data of the $19$ MS participants, whose symptoms were in line with CD1 or CD2. To further increase robustness of our models, each of these $19$ symptomatic participants was paired with a COVID-like symptom-free control participant with MS matched for site and gender and being at a similar age (cross-validation set). Thus, in each of the $19$ cross-validation rounds, the pre-trained models were further trained on the data of $18$ MS participants with COVID-like symptoms and $18$ MS participants without COVID-like symptoms, and subsequently tested on the data of the left-out MS participants pair. Table~\ref{fig:dp} reveals the numbers of participants per data subset as a function of the independent variables gender, age, and recording site. 

Heart rate data of the participants were assigned into temporal segments, defining a 14-day interval extending from 7 days preceding symptom onset to 7 days following symptom onset in which we sought to identify infection-related variations in heart rate. The interval mainly cover the duration of the COVID-19 incubation period \cite{Jantien2020,lauer2020,use2021}, and minimises the anomalous effects of day-to-day variations in activity, such as those observed between weekdays and weekends.

Fig.~\ref{fig:dp}
demonstrates the segmentation and subsequent data pre-processing procedure for the heart rate data of a participant with reported COVID-like symptoms. A heart rate segment over 14 days centred at 00:00 at the day of reported symptom onset, i.\,e., 7 consecutive days before the day of reported symptom onset plus 7 consecutive days starting with the day of reported symptom onset (red box on top of Fig.~\ref{fig:dp}) is referred to as \emph{symptomatic segment}. In contrast, an \emph{asymptomatic segment} stands for any 14-days interval of consecutive heart rate data again starting at 0 o'clock that is at least 7 days distant from a symptomatic segment (green box in top of Fig.~\ref{fig:dp}).  

 Thus, asymptomatic segments were created by shifting a 14-days window in full day steps over periods at least 7 days distant from the boundaries of a symptomatic segment. With the chosen 7-days distance of asymptomatic segments from symptomatic segments we presumed, that (i) a participant might not have already been infected 14 or more days prior to the onset of symptoms, and (ii) participants might have recovered from illness 14 days after the onset of symptoms at the latest. From the 49 participants of the pre-training set, totally 49 symptomatic segments and 1\,470 asymptomatic segments were extracted. Since the number of available symptomatic and asymptomatic segments was highly imbalanced, we replicated the symptomatic segments to the number of asymptomatic segments to guide the detection model weighted in favour of the minority class. For the LOSO CV procedure, 19 symptomatic and 570 asymptomatic segments were acquired from participants with reported symptoms, and 1\,140 asymptomatic segments from the control participants also referred to as \emph{control segments}. An overview of available symptomatic and asymptomatic segments is given in Table~\ref{tab:instances}.
 
\begin{table}[t!]
\caption{Available symptomatic and asymptomatic segments per data subset. Data completeness [\%] in terms of each heart rate segments is given in parentheses (mean + std).}
\label{tab:instances}
  \centering
  \setlength{\tabcolsep}{1mm}
  \setlength{\arrayrulewidth}{0.6pt}
  \renewcommand{\arraystretch}{0.9}
\begin{tabular}{l|ccc}
 \multicolumn{1}{l}{} & \multicolumn{1}{c}{} & \multicolumn{1}{c}{\textbf{Positive Participants}} & \multicolumn{1}{c}{\textbf{Health Control}}\\
 \multicolumn{1}{l}{\# (\%)} & \multicolumn{1}{c}{\textbf{Pre-training}} & \multicolumn{1}{c}{\textbf{for testing}} & \multicolumn{1}{c}{\textbf{for testing}}\\
 \hline
 \textbf{Symptomatic} &  $49$ $(98.7 \pm 0.3)$ & $19$ $(97.6 \pm 0.2)$  & $-$ \\
 \textbf{Asymptomatic}  &  $1470$ $(98.1 \pm 0.4)$ & $570$ $(97.4 \pm 0.2)$ & $1140$ $(99.2 \pm 0.5)$\\
 \hline
\end{tabular}
\vspace{-0.3cm}
\end{table}

Ideally, the instantaneous heart rate measurement for an individual was captured every second, and was uploaded every five seconds along with the recording time (blue curve in middle of Fig.~\ref{fig:dp}). 
The mean of 5 minutes is taken to smooth heart rate measurements while still 
tracking slow short-term changes in the heart rate. Moreover, this approach alleviates the effect of different sampling rates observed in actual heart rate recordings. 

Missing data over full 5-minutes intervals is filled with the median value of the overall 14-days segment. Finally, we have a single heart rate value per every 5 minutes. The resulting smoothed heart rate trajectory is considered appropriate to detect global heart rate patterns associated with COVID-like symptoms (red curve in middle of Fig.~\ref{fig:dp}).

\begin{landscape}
\begin{figure*}[th!]
  \centering
  \vspace{-1cm}
  \includegraphics[scale=0.45]{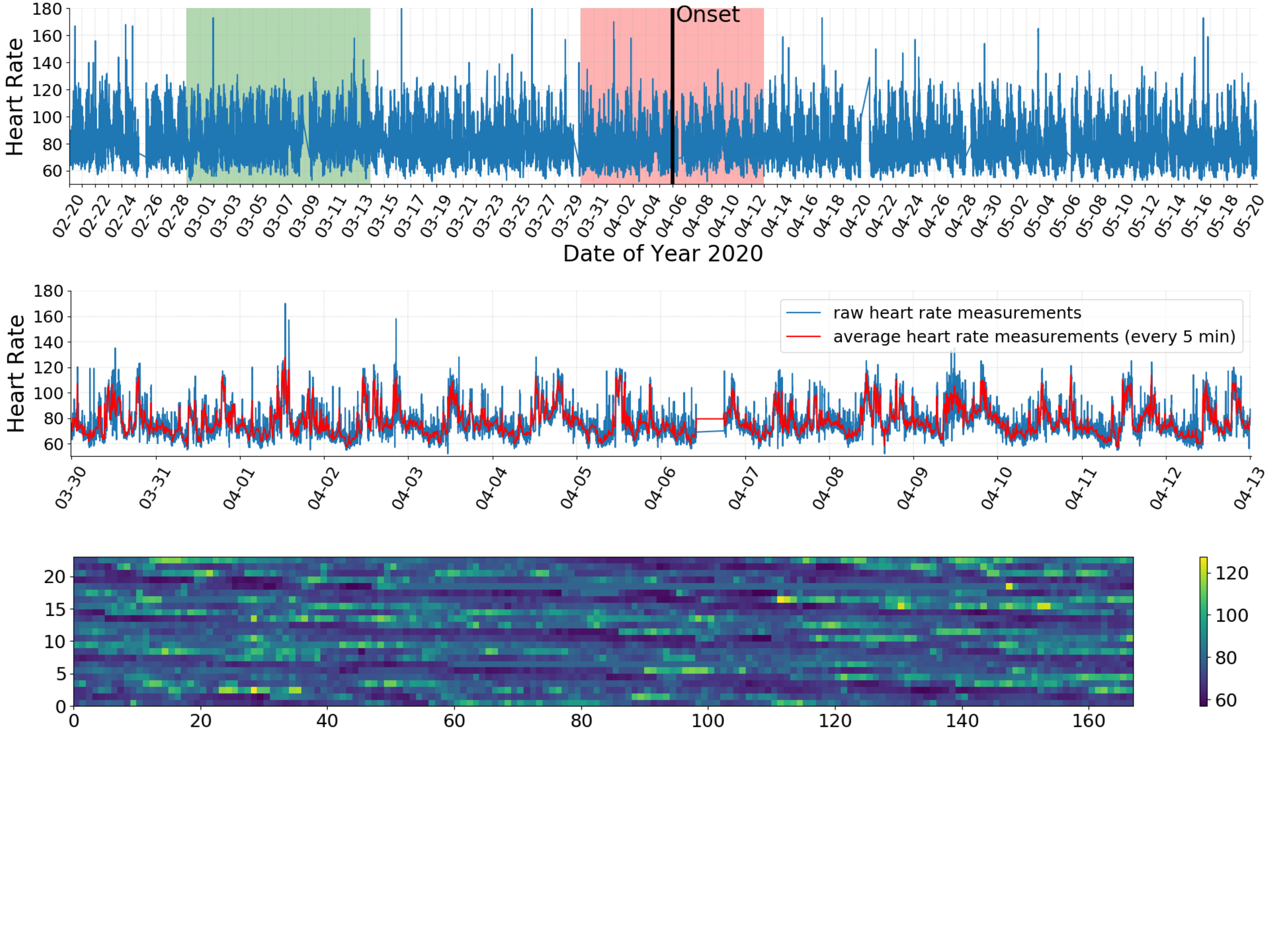}
  \vspace{-3cm}
  \caption{Segmentation and pre-processing of heart rate data of a participant with reported COVID-like symptoms. \textbf{Top}: Heart rate data recorded 24-hours-a-day/7-days-a-week from 21 February to 20 May 2020 (total 90 days). \emph{Onset} (black vertical bar) indicates 0 o'clock at the reported symptom onset date. The red rectangle covers 7 days heart rate data before and after symptom onset and represents a symptomatic segment, while the green rectangle gives an example of an asymptomatic segment, i.\,e., a segment in which no symptoms were present and which is located at least 7 days distant from the symptomatic segment. \textbf{Middle}: Symptomatic segment. The blue curve represents the unprocessed heart rate trajectory of the red rectangle above, while the red curve indicates the heart rate trajectory averaged over 5-minutes intervals. \textbf{Bottom}: Representation of the above illustrated symptomatic segment as $24 \times 168$ sized image of 5-minutes heart rate data related pixels. Each column represents an interval of 2 hours, the 168 columns sum up to 14 days.}
  \label{fig:dp}
\end{figure*}
\clearpage
\end{landscape}

We then transform the averaged heart rate data of each segment into a feature map, i.\,e., an image of size $24 \times 168$ pixels (bottom of Fig.~\ref{fig:dp}), in which each pixel represents a 5-minutes heart rate sampling point. Thereby, each column encodes a heart rate trajectory of 2 hours ($24 \times 5$ minutes), resulting in a covered interval of overall 14 days by 168 columns ($168 \times 2$ hours). In our experiments, we verified that this set-up of the feature map is effective as the input of our deep learning models, leading to promising detection results.

\section{Methodology}
\label{sec:method}


For detecting the presence of COVID-like symptoms in a given 14 days heart rate segment,
we investigate several different deep learning methods 
to represent the problem by feature maps, including CNNs \cite{olga2014,Alex2012} and a conventional CAE \cite{makkie2019fast,cheng2018,chen20172,du2017}. 
Further, we formulise the task as anomaly detection, and propose a strategy for training a CAE, by means of fitting the reconstruction error into the format of contrastive loss instead of conventional loss like root mean square error (RMSE). 

A CNN typically consists of a sequence of convolutional layers, with every convolutional layer containing its own learnable parameters, i.\,e., the filter weights and biases, to extract features from its input feature map. The stacking of convolutional layers allows a hierarchical decomposition of the input, a deeper CNN is expected to develop more complex representations of the the feature maps \cite{karen2014,zeiler2014,sakshi2018}. 
Subsequently, fully-connected layers map the learnt representations to the desired classes. Such CNN architectures have been successfully applied in many tasks, such as image and video recognition \cite{zhao2019,karpathy2014}, sequential data processing \cite{ismail2019,khamparia2019}, and medical applications \cite{yasaka2017,varun2016,bejnordi2017,olaf2015} including those recent works for Covid-19 diagnosis \cite{HU2018134,shi2020,oulefki2020automatic,bai2020}. 

The alternative approach to learn representations from an input image is to use a CAE \cite{ding2017,chen20172,du2017}. An auto-encoder contains two parts, i.\,e., an encoder and a decoder. The encoder learns latent attributes of the original input, whereas the decoder aims at reconstructing the original input from the learnt latent attribute. The dimensionality of the latent attributes is designed as a bottleneck imposed in the architecture; it hence can be seen as a compressed knowledge representation of the input. In order to guarantee the reproduction of the original input at the output of the decoder, the reconstruction error is minimised when optimising an auto-encoder network. In this work, we keep the architecture of our CAE encoder as consistent with our CNNs, and the decoder is arranged in a mirror-like fashion. Since the conventional CAE is an unsupervised learning approach, an additional multi-layer perception (MLP) \cite{gardner1998artificial} follows to classify the learnt attributes. 
This strategy has been widely used for medical and health related tasks \cite{chen2017,ruiz2017,Hu2018PredictionOD}, including tasks similar to ours, such as anomaly detection \cite{zhou2017anomaly,zong2018deep,zhao2017spatio,aytekin2018clustering}. 

As the strategy for optimising a conventional CAE does not take class information into account, the learnt latent attributes may contain redundant representations that are of no sufficient importance for the final classification. To incorporate the class information during training of the CAE, we apply contrastive loss \cite{khosla2020supervised} to the CAE reconstruction error in order to guide it to learn sufficiently discriminative latent attributes for different classes.
The three considered neural network architectures, i.\,e., CNN, conventional CAE, contrastive CAE, used in this work are described in details in the following separate sections.

\subsection{Convolutional Neural Network (CNN)}
\label{subsec:cnn}

\begin{figure}[t]
  \includegraphics[scale=0.52]{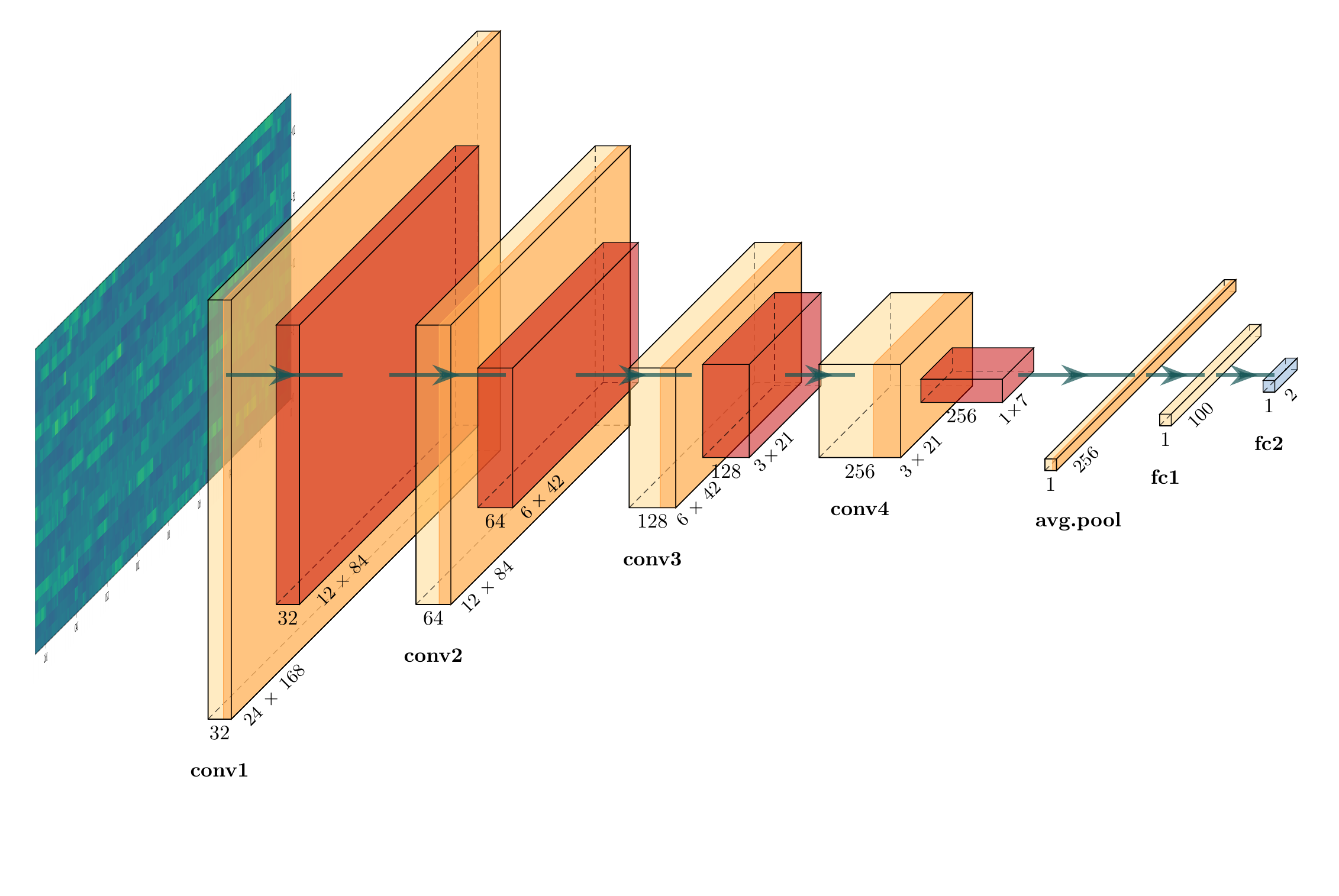}
  \vspace{-1.2cm}
  \caption{Convolutional neural network (CNN) architecture exemplified with 4 layers. Each layer involves a sequence of \textbf{convolution (conv)} -- \textbf{batch normalisation} -- \textbf{parametric rectified linear unit (PReLU) activation} -- \textbf{max pooling}. The output feature maps are averaged across all locations (avg.\ pool), and subsequently passed through two fully-connected (fc) layers to finally discriminate between two classes.}
  \label{fig:cnn}
\end{figure}

Our CNN architecture is comprised of a sequence of convolutional layers, an example of 4 layers is illustrated in Fig.~\ref{fig:cnn}. Each convolutional layer contains its own kernel size, stride, and number of channels as given in the specifications that appear in Table~\ref{tab:cnn}. For one layer convolution, kernel size and stride together determine the receptive field, i.\,e., the perception scope on the original input. Zero-padding is typically utilised to maintain a proper size of the feature map. More kernel filters result in better representation diversity. 
In between convoutional layers, batch normalisation can be used reduce the effect of internal covariate shift (ICS), an effect caused by different batches of training data having slightly different distributions \cite{bn2015,Bjorck2018}. Hence, the vanishing gradient issue can be alleviated. A parametric rectified linear unit (PReLU) \cite{he2015} performs as the activation function for each layer. It carries on the merits of ReLU activation \cite{shang2016understanding} in stable training while avoiding dead neurons using a learnable slope parameter.

\begin{table}[th]
  \caption{Specifications of the single convolutional (conv), pooling (pool), and fully-connected (fc) layers of the implemented convolutional neural network (CNN) model in terms of kernel size, stride size, padding size, and number of (\#) channels. avg.\ = average; * = dimensionality equals to the channel of the last convolutional layer.}
  \label{tab:cnn}
  \centering
  \setlength{\tabcolsep}{1.2mm}
  \setlength{\arrayrulewidth}{0.6pt}
  \renewcommand{\arraystretch}{0.9}
  \begin{tabular}{r | c c c r }
    \multicolumn{1}{r}{\textbf{Block}} & 
                          \multicolumn{1}{c}{\textbf{Kernel}} & \multicolumn{1}{c}{\textbf{Stride}} &
                          \multicolumn{1}{c}{\textbf{Padding}} &
                          \multicolumn{1}{c}{\textbf{\#\,Channels}} \\
    \hline
     \textbf{conv1} & $(5,5)$ & $(1,1)$ & $(2,2)$   & $32$ \\
     \textbf{pool1} & $(2,2)$ & $(2,2)$ & $-$       & $32$ \\
     \textbf{conv2} & $(5,5)$ & $(1,1)$ & $(2,2)$   & $64$ \\
     \textbf{pool2} & $(2,2)$ & $(2,2)$ & $-$       & $64$ \\
     \textbf{conv3} & $(5,5)$ & $(1,1)$ & $(2,2)$   & $128$ \\
     \textbf{pool3} & $(2,2)$ & $(2,2)$ & $-$       & $128$ \\
     \textbf{conv4} & $(5,5)$ & $(1,1)$ & $(2,2)$   & $256$ \\
     \textbf{pool4} & $(3,3)$ & $(3,3)$ & $-$       & $256$ \\
     \textbf{conv5} & $(3,3)$ & $(1,1)$ & $(1,1)$   & $512$ \\
     \textbf{conv6} & $(3,3)$ & $(1,1)$ & $(1,1)$  & $1024$ \\
     \hline
     \multicolumn{1}{c}{} & \multicolumn{4}{c}{\textbf{\# Hidden units}}\\
     \hline
     \textbf{avg. pool} & \multicolumn{4}{c}{$*$} \\
     \textbf{fc1}       & \multicolumn{4}{c}{$100$} \\
     \textbf{fc2}       & \multicolumn{4}{c}{$2$}    \\
    \hline
  \end{tabular}
  \vspace{-0.3cm}
\end{table}
A max-pooling layer processes the activations to reduce the spatial size of the feature size, leading to less parameters in the following network layers and, therefore, alleviating the potential over-fitting issues. The CNN is therefore a sequential cascade of convolutional layer -- batch normalisation -- PReLU. 
Its output is averaged across all locations, resulting in a representation vector. The representation vector is then projected onto the two classes, symptomatic or asymptomatic, through two fully-connected layers.

In our experiments, we tested CNNs with different number of layers, ranging from $1$ to $6$, in order to optimise the performance.


\subsection{Conventional Convolutional Auto-encoder (CAE)}
The encoder part of our CAE keeps the identical structure as the CNN introduced in Section \ref{subsec:cnn}, but contains only one FC layer before mapping into latent attributes. Given the feature map of a heart rate segment $x_i$, in which $i$ indicates segment index, the encoder $f^\textbf{enc}(\cdot)$ processes the feature map and produces the latent attributes
\begin{equation}
    h = f^\textbf{enc}(x_i).
\end{equation}

\begin{figure*}[t]
  \includegraphics[scale=0.35]{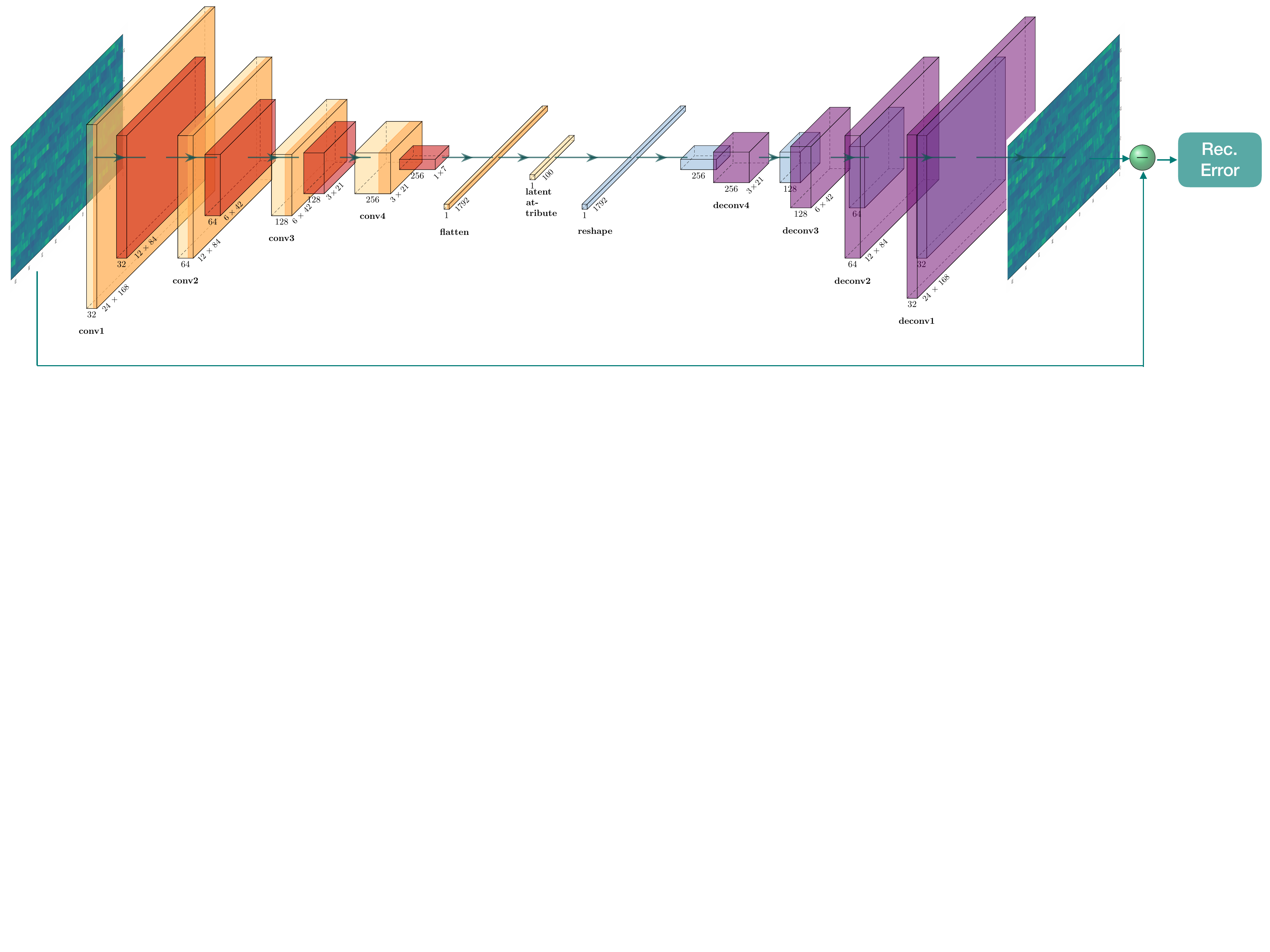}
  \vspace{-6.3cm}
  \caption{The convolutional auto-encoder (CAE) architecture with 4 encoder layers and 4 decoder layers as an example. Each encoder layer shares the same architecture as for the CNN in Fig.~\ref{fig:cnn} and the output feature maps are flattened and projected onto the latent attribute. In the decoder, the latent attribute is mapped back to the previous dimensionality. A decoder layer is a sequence of \textbf{transposed convolution} -- \textbf{batch-normalisation} -- \textbf{PReLU} -- \textbf{transposed max-pooling}. The output of decoder is seen as reconstructed image of the original input, and hence, the distance between the original and reconstructed image represents the reconstruction error.}
  \label{fig:ae}
\end{figure*}

The decoder presents an inverse processing of the encoder as seen in \ref{fig:ae}. For each decoder layer, the input feature map mainly passes through transposed convolution and transposed max-pooling, also known as de-convolution and de-pooling. Batch normalisation is employed in between, followed by PReLU as the activation function. The decoding procedure can be represented as 
\begin{equation}
    \hat{x}_i = f^\textbf{dec}(h),
\end{equation}
where $f^\textbf{dec}(\cdot)$ represent the operations in the decoder. Its output $\hat{x}_i$, seen as the reconstructed feature map, is then compared to the original input to compute the reconstruction error. 
An auto-encoder is optimised by minimising the reconstruction error, one typical such error is root mean squared error (RMSE) \cite{kenney1939mathematics}:
\begin{equation}
    \textbf{{RMSE}} = \sqrt{\frac{1}{N}\sum_i^N|x_i - \hat{x}_i|^2},
\end{equation}
where $N$ stands for the number of samples.

\begin{table}[th]
  \caption{The specifications of our CAE models. Each convolution and pooling layer, as well as de-convolution and de-pooling layer contains its own kernel size, stride, padding size, and number of channels. *=dimensionality depends on the total layers, **= dimensionality of latent attributes.}
  \label{tab:ae}
  \centering
  \setlength{\tabcolsep}{1.2mm}
  \setlength{\arrayrulewidth}{0.5pt}
  \renewcommand{\arraystretch}{0.7}
  \begin{tabular}{l | l | c c c r}
    \multicolumn{1}{c}{} &
    \multicolumn{1}{c}{\textbf{Blocks}} & 
                         \multicolumn{1}{c}{\textbf{Kernel}} & \multicolumn{1}{c}{\textbf{Stride}} &
                         \multicolumn{1}{c}{\textbf{Padding}} & \multicolumn{1}{c}{\textbf{\#\,Channels}} \\
                         
    \hline
     &\textbf{conv1} & $(5,5)$ & $(1,1)$ & $(2,2)$   & $32$  \\
     &\textbf{pool1} & $(2,2)$ & $(2,2)$ & $-$       & $32$  \\
     &\textbf{conv2} & $(5,5)$ & $(1,1)$ & $(2,2)$   & $64$  \\
     &\textbf{pool2} & $(2,2)$ & $(2,2)$ & $-$       & $64$  \\
     \textbf{CNN}&\textbf{conv3} & $(5,5)$ & $(1,1)$ & $(2,2)$   & $128$ \\
     \textbf{Encoder}&\textbf{pool3} & $(2,2)$ & $(2,2)$ & $-$       & $128$ \\
     &\textbf{conv4} & $(5,5)$ & $(1,1)$ & $(2,2)$   & $256$ \\
     &\textbf{pool4} & $(3,3)$ & $(3,3)$ & $-$       & $256$ \\
     &\textbf{conv5} & $(3,3)$ & $(1,1)$ & $(1,1)$   & $512$ \\
     &\textbf{conv6} & $(3,3)$ & $(1,1)$ & $(1,1)$  & $1024$ \\
     \cdashline{2-6}
     &\textbf{flatten} &  \multicolumn{4}{c}{$*$}    \\
     &\textbf{fc}    &   \multicolumn{4}{c}{$**$}    \\
    \hline
     &\textbf{fc}      &  \multicolumn{4}{c}{$**$} \\
     \cdashline{2-6}
     &\textbf{deconv6} & $(3,3)$ & $(1,1)$ & $(1,1)$  & $512$ \\
     &\textbf{deconv5} & $(3,3)$ & $(1,1)$ & $(1,1)$  & $256$ \\
     &\textbf{deconv4} & $(3,3)$ & $(1,1)$ & $(1,1)$  & $128$ \\
     &\textbf{unpool3} & $(3,3)$ & $(3,3)$ & $-$      & $128$ \\
     \textbf{CNN} &\textbf{deconv4} & $(5,5)$ & $(1,1)$ & $(2,2)$  & $64$ \\
     \textbf{Decoder}&\textbf{unpool4} & $(2,2)$ & $(2,2)$ & $-$      & $64$ \\
     &\textbf{deconv5} & $(5,5)$ & $(1,1)$ & $(2,2)$  & $32$ \\
     &\textbf{unpool5} & $(2,2)$ & $(2,2)$ & $-$      & $32$ \\
     &\textbf{deconv6} & $(5,5)$ & $(1,1)$ & $(2,2)$  & $1$ \\
     &\textbf{unpool6} & $(2,2)$ & $(2,2)$ & $-$      & $1$ \\
     \hline
  \end{tabular}
\end{table}

The specifications of our CAE are given in Table~\ref{tab:ae}. In experiments, we considered different number of convolutional layers in the CAE; the dimensionality of the flatten layer is determined by the last encoder layer. Hence, for different numbers of layers, we adjust the length of the fully-connected layer to optimise the CAE performance.

\subsection{Contrastive CAE}


The difficulty in finding good latent attributes in the framework of an  auto-encoder lies in setting it to a proper dimensionality. Too long latent attributes may contain redundancies for easier reconstructing the original input, but fall short of concentrating on learning the saliently discriminative features for different classes. Meanwhile, shorter latent attributes can have less or limited representation capability. 
The conventional optimisation strategy for an auto-encoder considers no class information \cite{chen20172,du2017}, and hence the learnt latent attributes is not well oriented to be discriminative for different classes. 
Specifically, for our detection task, the auto-encoder may tend to learn the latent attributes that can better reconstruct the original feature map, while ignoring some salient attributes that indicate the difference between symptomatic and asymptomatic segments.  

To incorporate class information, symptomatic and asymptomatic, into the optimisation of CAE, we fit the reconstruction error of the two classes into contrastive loss\,\cite{khosla2020supervised}. As analogues to anomaly detection, we expect the CAE to output a low reconstruction error for asymptomatic segments, and a high reconstruction error for symptomatic segments.
Therefore, the loss function for our contrastive CAE can be seen as 
\begin{equation}
    \begin{split}
        \textbf{Contrastive Loss} = & \sqrt{\frac{1}{N}\sum_i^N|x_i^p - \hat{x}^{p}_i|^2} + \\ & (\textbf{m} - \sqrt{\frac{1}{N}\sum_i^N|x_i^n - \hat{x}^{n}_i|^2}),
    \end{split}
\label{eq:cl}
\end{equation}
where the superscripts $p$ and $n$ are used to distinguish positive (symptomatic) and negative (asymptomatic) samples. 
Ideally, the reconstruction error for a positive pair, original and reconstructed feature map for asymptomatic segments, is expected to be $0$, indicating a successful reconstruction of the original input at the decoder. In contrast, the reconstruction error for the negative input pair, original and reconstructed feature map for symptomatic segments, is expected to be close to the margin value $m$. 
Therefore, the difference in classes leads to different reconstruction errors from our CAE.

Contrastive loss has been successfully exploited in several image processing tasks, especially those recently proposed for CT and X-ray diagnosis of COVID-19\,\cite{,SHORFUZZAMAN2021107700,LI2021107848,CHEN2021107826}. 




\section{Experiments \& Results}
We conducted a series of experiments in order to test the models presented in Section III for detecting COVID-like symptoms in the given 14-days heart rate segments. The neural networks of different architectures are pre-trained with the heart rate segments of 49 participants that reported COVID-like symptoms, but had not met the CD1 or CD2 criterion. We apply leave-one-subject-out cross-validation to the heart rate segments of the 19 individuals who meet CD1 or CD2, and further test the models on the corresponding symptom-free control group.

The performance of our most effective method, i.\,e., the Contrastive CAE ($m$ is set to $5$), is mainly compared to the conventional CNN and conventional CAE that is optimised with an RMSE loss. Models of different layers are tested using average Unweighted Average Recall (UAR, \ie, the recall of the classes added divided by the number of classes), Precision, F1 measure, sensitivity, and specificity as the evaluation metrics throughout the experiments.

To explore the advantage of using contrastive loss instead of RMSE loss when training an CAE, we consider the dimensionality of latent attributes in different lengths, including 50, 100, 300, 500, and 1\,000, ensuring fair comparisons. 
A two-layers MLP is separately optimised for each model of different layers to project the learnt latent attributes to classes, symptomatic or asymptomatic. 
Furthermore, since the contrastive CAE is expected to produce a low reconstruction error for positive pairs (asymptomatic), and a high reconstruction error for negative pairs (symptomatic), it provides us the possibility to directly perform a classification based on the reconstruction error using classic machine learning techniques, for instance, logistic regression. 

The models' parameters are optimised using an Adam optimiser. The learning rate decays from $0.03$ to about $0.0001$ with a decay factor of $0.33$ after every $50$ epochs. We keep using a batch size of $32$ for all experiments. The settings of the hyper-parameters are selected after careful fine-tuning the experiments, to assure stable and fast convergence of our models. 

In the following, we first compare our proposed contrastive CAE with a CNN model, both of which use a different number of layers. As an additional comparison method, we applied an MLP directly to the one-dimensional 5-min average heart rate segment without formatting it into feature maps.

\subsection{Contrastive CAE \textbf{vs} MLP \textbf{vs} CNN}
\begin{table}[t]
  \caption{Evaluation results for the binary COVID-19 yes/no (based on the symptom CD1/CD2 definitions above) classification [\%] of the CNN and Contrastive CAE models with a different number of layers. For the Contrastive CAE, classification is performed based on reconstruction error using logistic regression. Unweighted Average Recall (UAR, chance-level is 50\,\%), Prescision, F1 measure, Sensitivity and Specificity are used as evaluation metrics.}
  \label{tab:lys}
  \centering
  \setlength{\tabcolsep}{0.6mm}
  \setlength{\arrayrulewidth}{0.6pt}
  \renewcommand{\arraystretch}{0.7}
  \begin{tabular}{l | c | ccccc}
    \multicolumn{1}{c}{} &
    \multicolumn{1}{c}{\rotatebox{30}{\textbf{\#\,Layers}}} & 
                         \multicolumn{1}{r}{\rotatebox{30}{\textbf{UAR}}} & \multicolumn{1}{r}{\rotatebox{30}{\textbf{Precision}}} & \multicolumn{1}{r}{\rotatebox{30}{\textbf{F1 measure}}} & \multicolumn{1}{r}{\rotatebox{30}{\textbf{Sensitivity}}} & \multicolumn{1}{r}{\rotatebox{30}{\textbf{Specificity}}}\\
    \hline
    \textbf{MLP} & & $61.0$ & $50.5$ & $38.6$ & $63.2$ & $58.8$\\
    \hline
     & $1$ & $49.7$ & $50.0$ & $62.8$ & $52.6$ & $46.8$\\
     & $2$ & $63.7$ & $71.9$ & $59.1$ & $63.2$ & $64.3$ \\
    \textbf{CNN} & $3$ & $\textbf{{76.0}}$ & $\textbf{{80.5}}$ & $\textbf{{75.5}}$ & $ \textbf{{79.0}}$ & $\textbf{{73.1}}$\\
     & $4$ & $66.4$ & $73.8$ & $62.7$ & $68.4$ & $64.3$\\
     & $5$ & $55.6$ & $50.3$ & $55.0$ & $52.6$& $58.5$\\
     & $6$ & $60.8$ & $50.5$ & $58.5$ & $63.2$ & $58.5$\\
    \hline
     & $1$ & $58.8$ & $49.2$ & $45.1$ & $70.2$& $47.4$ \\
     & $2$ & $83.0$ & $69.7$ & $74.5$ & $84.2$ & $81.9$ \\
     \textbf{Contrastive} & $3$ & $90.6$ & $84.0$ & $84.6$ & $\textbf{{100.0}}$ & $81.3$\\
     \textbf{CAE} & $4$ & $\textbf{{95.3}}$ & $86.4$ & $\textbf{{88.2}}$ & $\textbf{{100.0}}$ & $\textbf{{90.6}}$\\
     & $5$ & $93.9$ & $\textbf{{86.7}}$ & $88.0$ & $\textbf{{100.0}}$ & $87.7$\\
     & $6$ & $90.9$ & $83.5$ & $84.6$ & $\textbf{{100.0}}$ & $81.9$\\
    \hline
  \end{tabular}
\end{table}

The MLP is found to be 
best be 
optimised with $4$ layers ($1000-250-50-20$ hidden units of each layer), and its performance is shown in Table~\ref{tab:lys}. In terms of all evaluation metrics, the CNN model achieves its best performance with $3$ convolutional layers, demonstrating significant improvements over the MLP baseline ($p<0.05$) in a one-tailed z-test.
Further significant improvement can be achieved by using the contrastive CAE.
We apply logistic regression to the reconstruction error of the test set, and observe that the contrastive CAE with 4 encoder and 4 decoder layers performs the best, achieving a UAR of $95.0\%$, precision of $86.4\%$, F1 measure of $88.2\%$, sensitivity of $100.0\%$, and specificity of $90.6\%$. Regarding the precision,
the contrastive CAE containing 5 encoder and 5 decoder layers can perform slightly better. The performance results in significant improvements over the CNN approaches ($p<0.05$) in a one-tailed z-test.


\subsection{Contrastive CAE \textbf{vs} Conventional CAE}
We next investigate training a convolutional auto-encoder using contrastive loss instead of RMSE for the binary COVID-19 (based on the symptoms described above) classification task. Different dimensionality of latent attribute were tested, using a two-layers MLP classifier to project latent attributes to classes. The parameters of the MLP classifier are separately tuned for different dimensionalities in order to optimise each models' detection performance.
The conventional CAE reaches its optimum with the latent attributes of the size of $50$, achieving $60.1\%$ UAR, $51.9\%$ precision, $50.5\%$ F1 measure, $50.5\%$ sensitivity, and $75.4\%$ specificity as given in Table~\ref{tab:comparison}. This indicates that the CAE trained with RMSE has a limited capability in learning discriminative lattent attributes between symptomatic and asymptomatic segments, since the attributes learning procedure considers no class information during optimisation, and, hence leaves the classification difficulty to  
the final MLP classifiers. 
Besides, the optimised performance of the conventional CAE is worse than that of the CNN models, further stressing the need of involving the class information in training a more efficient CAE for the detection task.

\begin{table}[th]
  \caption{Comparison of the results [\%] between convolutional auto-encoders with 4 encoder and 4 decoder layers trained with RMSE loss vs contrastive loss. Classification are performed based on the latent latent attributes. Binary COVID-19 yes/no (based on symptom CD1/CD2 definitions above classification.)
  \#\,Attr: dimensionality of latent attributes.}
  \label{tab:comparison}
  \centering
  \setlength{\tabcolsep}{0.6mm}
  \setlength{\arrayrulewidth}{0.6pt}
  \renewcommand{\arraystretch}{0.7}
  \begin{tabular}{l | r | c c c c c}
     \multicolumn{1}{c}{} & \multicolumn{1}{c}{\rotatebox{30}{\textbf{\#\,Attr}}} &  \rotatebox{30}{\textbf{UAR}} & \rotatebox{30}{\textbf{Precision}} & 
     \rotatebox{30}{\textbf{F1 measure}} &   \rotatebox{30}{\textbf{Sensitivity}} & \rotatebox{30}{\textbf{Specificity}}\\
    \hline
     & $50$ & $\textbf{{66.6}}$ & $\textbf{{51.9}}$ & $\textbf{{50.5}}$ & $57.9$ & $75.4$\\
     & $100$ & $58.5$ & $51.2$  & $48.7$ & $47.4$ & $69.5$\\
     \textbf{CAE} & $300$ & $63.4$ & $51.0$ & $44.6$ & $63.2$ & $63.7$\\
     \textbf{} & $500$ & $65.8$ & $51.3$ & $49.1$ & $68.4$ &$63.2$\\
     & $1000$ & $55.3$ & $50.5$ & $37.5$ & $47.4$ & $63.2$\\
    \hline
     & $50$ & $92.0$ & $64.3$ & $70.4$ & $100.0$ & $83.9$\\
     \multirow{2}{*}{\textbf{Contrastive}}& $100$ & $\textbf{{92.2}}$ & $67.2$ & $73.9$ & $100.0$ & $84.3$\\
     \multirow{2}{*}{\textbf{CAE}}& $300$ & $90.9$ & $\textbf{{84.0}}$ & $\textbf{{84.6}}$ & $\textbf{100.0}$ & $81.9$\\
      & $500$ & $90.9$ & $64.4$ & $70.3$ & $94.7$ & $\textbf{87.1}$\\
     & $1000$ & $71.9$ & $53.3$ & $53.1$ & $68.4$ & $75.4$ \\
     \hline
  \end{tabular}
\end{table}


For the bianry classification task, the classes' difference can be implicitly modelled in the contrastive loss as in Eq. (\ref{eq:cl}) for training the CAE, since the positive and negative reconstruction error are guided to produce a margin between each other in a discriminative manner. Hence, the contrastive CAE is capable of learning latent attributes that represent salient features to distinguish between symptomatic and asymptomatic segments. Again, a 2-layer MLP is used to classify the learnt latent attributes into classes, symptomatic or asymptomaic.
In our experiments, the contrastive CAE with an attribute dimensionality of $100$ achieves its best result in terms of UAR, and when the dimensionality increases to $300$, the proposed contrastive CAE achieves a balanced results between the UAR and precision measure, leading to its best F1 measure, i.\,e., $84.6\%$. In addition, the model obtained the optimum sensitivity and a high specificity in the detection task, which considerably  outperforms the conventional CAE that is trained with RMSE loss.

\begin{table}[th]
  \caption{The classification results [\%] of contrastive convolutional auto-encoders with 4 encoder and 4 decoder layers based on the reconstruction error (rec. error) using logistic regression. Binary COVID-19 yes/no (based on symptom CD1/CD2 definitions above classification. The last row indicates removing the latent attributes layer.}
  \label{tab:attrlen}
  \centering
  \setlength{\tabcolsep}{0.6mm}
  \setlength{\arrayrulewidth}{0.6pt}
  \renewcommand{\arraystretch}{0.7}
  \begin{tabular}{l | r | c c c c c}
     \multicolumn{1}{c}{} & \multicolumn{1}{c}{\rotatebox{30}{\textbf{\#\,Attr.}}} &  \rotatebox{30}{\textbf{UAR}} & \rotatebox{30}{\textbf{Precision}} & 
     \rotatebox{30}{\textbf{F1 measure}} &   \rotatebox{30}{\textbf{Sensitivity}} & \rotatebox{30}{\textbf{Specificity}}\\
    \hline
     & $50$ & $93.9$ & $83.9$ & $87.5$ & $100.0$ & $87.7$\\
     \textbf{Contrastive} & $100$ & $\textbf{{95.3}}$ & $\textbf{{86.4}}$ & $\textbf{{88.2}}$ & $\textbf{{100.0}}$ & $\textbf{{90.6}}$\\
     \textbf{CAE} & $300$ & $91.5$ & $77.9$ & $84.7$ & $100.0$ & $83.0$\\
     \textbf{(rec. error)} & $500$ & $92.4$ & $80.6$ & $85.4$ & $100.0$ & $84.8$\\
     & $1000$ & $94.4$ & $81.7$ & $88.1$ & $100.0$ & $88.9$\\
     & $-$ & $93.3$ & $71.6$ & $76.8$ & $100.0$ & $86.6$\\
     \hline
  \end{tabular}
\end{table}

Since our contrastive CAE was trained to create a sufficient margin between the reconstruction errors of symptomatic and asymptomatic segments, applying classification directly on the reconstruction errors, rather than the learnt latent attributes, is a more efficient way to use the contrastive CAE for our binary decision task. 
The decision threshold between the reconstruction errors of symptomatic and asymptomatic classes is determined by the means of applying logistic regression \cite{Park2013AnIT} on the training part for each cross-validation round. A 14-days heart rate segment is decided for as COVID-19 symptomatic (CD1/CD2 criterion) if the reconstruction error is above the decision boundary. 

The best performance, shown in Table~\ref{tab:attrlen}, is achieved with the attributes' length equalling $100$, achieving $95.0\%$ UAR, $86.4\%$ precision, $88.2\%$ F1 measure, $100.0\%$ sensitivity, and $90.6\%$ specificity.
Generally, the contrastive CAE performs stable over different attributes' dimensionality, reducing the difficulty in setting its proper dimensionality. An extreme case is to combine the encoder and decoder by removing the latent attributes layer. The performance, however, maintains stable as given in the last row of Table~\ref{tab:attrlen}. 

\subsection{Continuous Classification}
\label{sec:conDet}
We further explore the possibility to make binary COVID-19 yes/no (based on the symptom CD1/CD2 definitions above) decisions earlier with our proposed contrastive CAE. To this end, we shift the window for sliding over the symptomatic segments to earlier and later days, but still contain the onset date. On the other hand, in order to test the model's validality on later days, we shift the sliding window for asymptomatic segments to later days that contain the onset date. In our experiment, we keep the asymptomatic segments as in the previous experiments, but replace the previous symptomatic segments with the shifted symptomatic segments.

The experimental results, as seen in Table~\ref{tab:detection}, reveal that the model works well for the heart rate segments that are sliced one day earlier or three days later, leading to no performance degradation. This is potentially due to the fact that the model was trained with segments centred by onset date, 
and the shifted symptomatic segments share similar internal structure as the original symptomatic segments. However, segments that further deviate from the original symptomatic segments, i.\,e., shifting the sliding window to two more previous days or four days later, results in decreased classification accuracy. 

\begin{table}[th]
  \caption{Test results [\%] for shifting the context window; binary COVID-19 yes/no (based on symptom CD1/CD2 definitions above) classification.}
  \label{tab:detection}
  \centering
  \setlength{\tabcolsep}{0.6mm}
  \setlength{\arrayrulewidth}{0.6pt}
  \renewcommand{\arraystretch}{0.7}
  \begin{tabular}{l | r | c c c c c}
    \multicolumn{1}{c}{} &
    \multicolumn{1}{c}{\rotatebox{30}{\textbf{\#\,Days}}} & 
                         \multicolumn{1}{c}{\rotatebox{30}{\textbf{UAR}}} & \multicolumn{1}{c}{\rotatebox{30}{\textbf{Precision}}} & \multicolumn{1}{c}{\rotatebox{30}{\textbf{F1 measure}}} &
                         \multicolumn{1}{c}{\rotatebox{30}{\textbf{Sensitivity}}} &
                         \multicolumn{1}{c}{\rotatebox{30}{\textbf{Specificity}}}\\
    \hline
     & $-3$ & $71.6$ & $52.7$ & $52.7$ & $52.6$ & $62.2$\\
     & $-2$ & $79.5$ & $53.6$ & $54.2$ & $68.4$ & $61.0$\\
     & $-1$ & $97.5$ & $89.2$ & $91.3$ & $100.0$ & $91.2$\\
     & $0$ & $\textbf{{95.3}}$ & $\textbf{{86.4}}$ & $\textbf{{88.2}}$ & $\textbf{{100.0}}$ & $\textbf{{90.6}}$\\
     \textbf{Contrastive}& $1$ & $96.7$ & $87.8$ & $89.8$ & $100.0$ & $90.8$\\
     \textbf{CAE} & $2$ & $96.1$ & $87.2$ & $89.1$ & $100.0$ & $90.3$\\
     & $3$ & $96.3$ & $86.5$ & $88.4$ & $100.0$ & $89.9$\\
     & $4$ & $81.3$ & $86.6$ & $81.8$ & $94.7$ & $80.2$\\
     & $5$ & $60.5$ & $60.8$ & $60.3$ & $68.4$ & $54.6$\\
    \hline
  \end{tabular}
\end{table}

\begin{figure*}[t!]
\centering
\hspace*{-0.5cm}
  \includegraphics[scale=0.36]{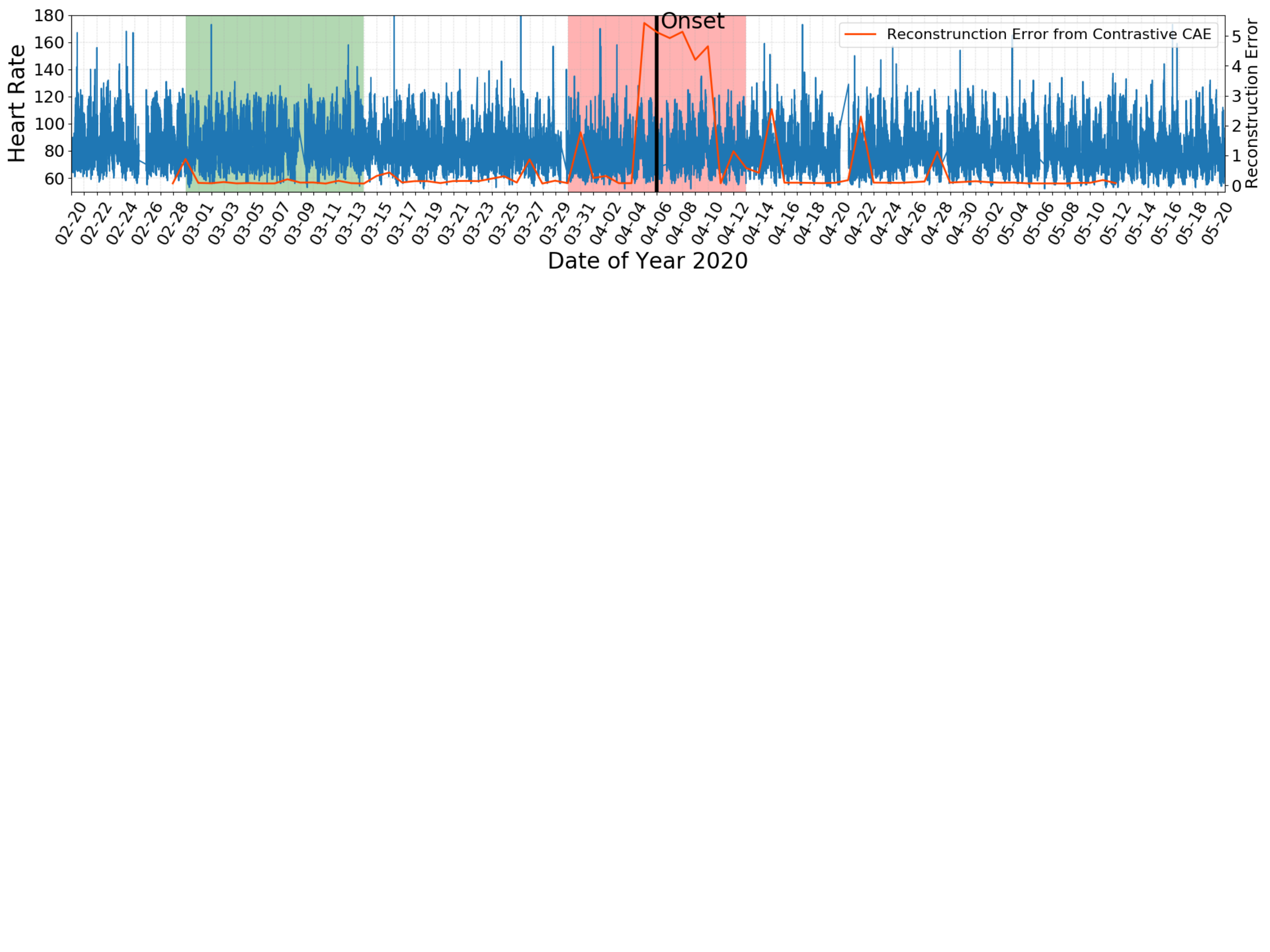}
   \vspace{-7.8cm}
  \caption{Continuous binary COVID-19 yes/no (based on the symptom CD1/CD2 definitions above) classification on each given 14-days heart rate windows of an examplary individual (the same as in Figure \ref{fig:dp}, top).}
  \label{fig:detection}
\end{figure*}

Fig.~\ref{fig:detection} illustrates using our contrastive CAE to continuously classify COVID-19 yes/no based on CD1/CD2 symptom presence based on the example given in \ref{fig:dp}, top. 
This explicitly shows that the reconstruction error for the contrastive CAE is low for the asymptomatic segments, but a decision can be made from the 14-days segments centred by date from one day before onset to three days after onset. This indicates that the onset detection of the COVID-like symptoms can be estimated in their earlier stage up to several days later.

\section{Discussion}
\subsection{Shifting of Symptomatic Segments}
Throughout all experiments, we kept assuming that the participant-reported onset date is identical to the real symptom onset. As seen in Section \ref{sec:conDet}, the contrastive CAE performs effective on the symptomatic segments that are centred relating to the reported symptom onset. In addition, we observe the CAE to work robustly on the segments one day earlier and three days later, and further shifting of the symptomatic segments leads to decrease in performance. 

However, considering that the participants may have hesitated before reporting the appearance of their symptoms, the true symptom onset may happen earlier than the reported onset date. 
As the experimental results given in Table~\ref{tab:detection}, we see that the proposed model starts giving good results one day earlier, although the model was optimised only with symptomatic segments and their control segments, indicating that potentially some participants were conscious of the symptoms and reported the symptoms on the second day. Furthermore, shifting the segments up to a few days later (maximally three days in our experiments) leads to higher  certainty that symptoms are indeed contained, and hence, our model achieves stable performance in this case.  
Potentially, there exist some participants whose true symptom started on earlier days, but relieved soon. In this case, the segments, that are shifted to a later point in time by many days, may exclude the true symptomatic part, leading to a low detection accuracy when assuming that the considered segments contain symptoms.

\subsection{Features from Heart Rate Measurements}
In this work, we base our study on 5-minutes mean of the heart rate measurements. In practice, we also considered the heart rate variance every 5 minutes as input features when starting with standard CNNs, and did not see additional improvement in terms of accuracy. Therefore, in further studies of conventional CAEs and contrastive CAEs, we concentrated on the mean value rather than variance. The mean heart rate values as features of our deep learning models may be sufficient, since the global variance is retained in segmental mean values. 

\subsection{Difficulty in Model Optimisation}
In some previous works, the class information can be applied to latent attribute layers, leading to the supervised auto-encoder introduced in \cite{NIPS2018_7296}. Cross-entropy losses are used to minimise the difference between predicted labels from latent attributes and true labels. This approach provides a certain preservation of the reconstructed feature map, taking the cross-entropy loss as a regularisation method. The reconstruction error and cross entropy loss are jointly optimised. However, the optimisation of the joint loss requires a proper combination factor in order to balance the optimisation on reconstruction error and prediction error. Since the two types of errors originate from different stages of the auto-encoder model, leading to their different scale level, the difficulty lies in seeking a good combination scale.

In our proposed method, the contrastive loss can be employed directly on the reconstruction error for positive and negative input pairs. The training procedure is simpler, since the reconstruction errors of the positive pairs and negative pairs present in the same scale level at the beginning of the training stage. 
The method also provides the chance to validate whether the model has learnt discriminative latent attributes for different classes in the auto-encoder framework.

\section{Conclusion}
In this work, we explored several deep learning models, including conventional CNNs, conventional CAEs, and contrastive CAEs, 
to make a machine-learning-based COVID-19 yes/no decision based on symptoms' presence as definded by two criteria (CD1/CD2) given 14-days heart rate measurements from a Fitbit wristband. The models were pre-trained based on the heart rate data of $49$ participants with MS who reported to have COVID-like symptoms but did not satisfy CD1 or CD2. However, the reported symptoms were highly related to COVID-19 symptoms. The models were then tested on data of $19$ MS participants whose reported symptoms met the criteria of CD1 or CD2, by means of LOSO CV. In this process, each of the $19$ symptomatic MS participants was paired with a site-, gender-, and age-matched symptom-free MS participants. Experimental results indicate that our proposed contrastive CAE approach, incorporating class information into optimising the CAE with contrastive loss, achieved considerable
improvements over the conventional CNN and CAE models in terms of performance, including UAR, precision, F1 measure, as well as sensitivity, and specificity.

The effectiveness of contrastive learning as demonstrated in this work shall motivate further research, including the investigation of different bio-signal measurements, features, and model architectures. Several recent works \cite{nature2020,Mishra2020} have shown the potential of using more than one bio-signal measurement for Covid-19 recognition. Such additional measurements can be audio and speech signal \cite{schuller2020,Jing2020} which can also be achieved from wearable devices.

Since the set-up of our experiments was chosen to detect whether or not the COVID-19-like symptoms appeared during a period of heart rate data, the models show limitations in a causal set-up, i.\,e., when trying to predict potential symptoms before they are present. To this end, future work shall try to answer the question of how many days in advance we will already be able to reliably predict the potential imminent onset of COVID-like symptoms. As the acquisition of data in the RADAR-CNS programme is still ongoing, the improvement of our proposed binary COVID-19 yes/no (based on the symptom CD1/CD2 definitions above) classification model based on a broader data foundation is expected. Further to that, other windows of time should be analysed. Overall, we are optimistic that an applicable decision can be made as to COVID-19 presence based on the symptoms defined herein based on machine learning analysis of consumer-type heart rate measurement.

\section*{Acknowledgement}
This project has received funding from the Innovative Medicines Initiative\footnote{\url{https://www.imi.europa.eu/}} 2 Joint Undertaking under grant agreement No 115902. This Joint Undertaking receives support from the European Union’s Horizon 2020 research and innovation programme and EFPIA. This communication reflects the views of the RADAR-CNS consortium and neither IMI nor the European Union and EFPIA are liable for any use that may be made of the information contained herein.






\bibliography{mybibfile}

\end{document}